\documentclass[a4paper,12pt]{article}
\usepackage[a4paper,inner=3cm,outer=2cm,top=4cm,bottom=2cm]{geometry}
\usepackage[scaled=0.91]{helvet}
\usepackage[english]{babel}
\usepackage[utf8x]{inputenc}
\usepackage[onehalfspacing]{setspace}
\usepackage{parskip}
\usepackage{amsmath}
\usepackage{graphicx}
\usepackage[colorinlistoftodos]{todonotes}
\usepackage{float}
\usepackage{hyperref}
\usepackage{datetime}
\newdate{date}{12}{06}{2018}
\date{\displaydate{date}}

\begin{document}

\begin{titlepage}

\newcommand{\HRule}{\rule{\linewidth}{0.5mm}} % Defines a new command for the horizontal lines, change thickness here

\center % Center everything on the page
 
%----------------------------------------------------------------------------------------
%	HEADING SECTIONS
%----------------------------------------------------------------------------------------

\textsc{\LARGE Bogazici University}\\[1.5cm] % Name of your university/college
\textsc{\Large CMPE 492}\\[0.5cm] % Major heading such as course name
\textsc{\large Senior Design Project}\\[0.5cm] % Minor heading such as course title

%----------------------------------------------------------------------------------------
%	TITLE SECTION
%----------------------------------------------------------------------------------------

\HRule \\[0.4cm]
{ \huge \bfseries Load Balancing and Mutisource Routing in Information-Centric-Networking }\\[0.4cm] % Title of your document
\HRule \\[1.0cm]
 
%----------------------------------------------------------------------------------------
%	AUTHOR SECTION
%----------------------------------------------------------------------------------------

\begin{minipage}{0.4\textwidth}
%\begin{flushleft} \large
\emph{Authors:}\\
Nihal Ezgi \textsc{YUCETURK} \\ % Your name
Kutay \textsc{CANDAN} % Your name
%\end{flushleft}
\end{minipage}
~
\begin{minipage}{0.5\textwidth}
%\begin{flushright} \large
\emph{Supervisors:} \\
Prof. Tuna \textsc{TUGCU} \\% Supervisor's Name
Asst. Prof. M. Sukru \textsc{KURAN} % Supervisor's Name
%\end{flushright}
\end{minipage}\\[2cm]

%\{\large \today}
\begin{center}
    {\large June, 2018}\\[2cm]
\end{center}% Date, change the \today to a set date if you want to be precise

\includegraphics{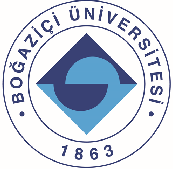}\\[1cm] % Include a department/university logo - this will require the graphicx package
 
%----------------------------------------------------------------------------------------

\vfill % Fill the rest of the page with whitespace

\end{titlepage}

\tableofcontents
\newpage
\begin{abstract}
\noindent Information-Centric Networking is still an incomplete paradigm which introduces a large variety of new topics and approaches over the traditional networking. Since it is a relatively new concept which promises for easier and faster data access many aspects of it have been studied intensely in recent years. Nonetheless, load balancing is one of the least focused but high potency, open-ended areas. The aim of this project is to present an alternative technique to load balancing in the domain of pre-suggested routing method LSCR (Link State Content Routing). We suggested multi-path interest and multi-source data packet routing on the bases of Forwarding and Routing Tables of LSCR and packet routing protocol of Source Routing. We have developed a simulator software to implement and test the multi-path routing algorithm. Multi-path routing and LSCR are compared with regard to the load of the links, package loss, package delay and overall network performance
\end{abstract}
\newpage

\section{Introduction and Motivation}

\subsection{Our Motivation to Work on ICN}
Today’s network architecture was designed many years ago (the 60s-70s) when there were only a few hosts and clients and when all their needs were to share limited and precious resources. Internet Protocol (IP) showed up and was implemented around the 80s, was naturally simple and convenient to use with different topologies and different technologies. The major concern of that time was where to forward to get intended content and how. Contents were relatively small and simple -mostly text files- and traffic was not heavy because of the limited number of users. Mobility was not an issue because users were not mobile and computers were too large to be moved around. Security was an afterthought problem because of fixed lines somewhere in the ground, not that easy to access. Thus, the overall system was able to meet the overall demand.

\begin{figure}[H]
\centering
\includegraphics[width=1.0\textwidth]{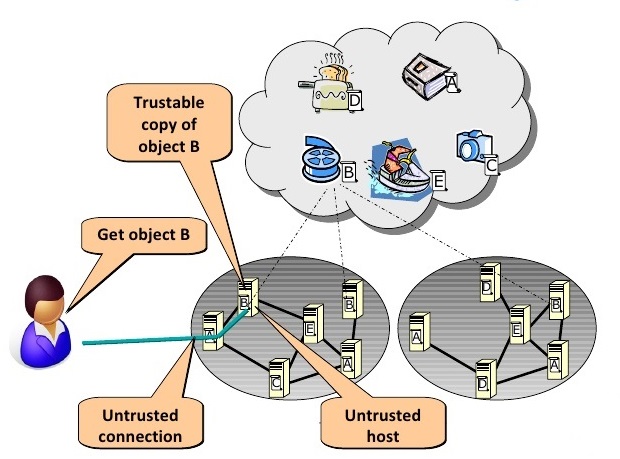}
\caption{\label{fig:ICNoverview}ICN Representation}
\end{figure}
\newpage
In today’s networking, we have multi-interfaced devices, wireless communication possibilities, various technologies to support mobility such as 4G,5G and high-speed links and processors at core devices as well as at end devices. Reaching to a specific host is almost guaranteed so what to get is a major concern rather than from where or how to get. Mobility and security are essential because now computers -phones, tablets etc.-are small enough to move around and even the most vital practices -bank transactions, elections- are performed via networks. The current architecture was not designed to meet these usages for sure, so we may want a new structure to support efficient data sharing, mobility and security.
\\[5pt]
ICN is an expanding paradigm and growing field of study. Since it may affect overall Internet where everything is somehow is depended and related. It is not only the concern of academia but also is of particular concern to big cooperations. We believe that even ICN architecture will be not in use soon, working on ICN has rewarding advantages. Among all sub-fields of ICN, routing was the most suitable and promising one when considering our level of study. Thus, we decided to work on the network and transport layers of the architecture.

\subsection{General Knowledge About ICN}
Information-Centric Networking changes the paradigm of the usual Internet which consists of hosts, clients, servers and routers. It turns named-host structure into named-content; clients and routers. The network itself looks for specific data rather than specific host which serves the data. The new design envisages caching mechanism which uses internal memory of routers to ensure a widespread of contents available in the network. The new design provides security for data rather than hosts or client.
\\[5pt]
ICN can be divided into five distinct subtopics which describe the whole system altogether.\cite{ICNSurvey2014} An operation performed by ICN consists of four different components which are data object, caches, routing and forwarding and overall security.[3] In an ICN network, clients send an interesting message to get a specific content and the network endeavours to provide the correct content from the nearest location which it resides. Content is identified by naming and name resolution mechanism whereas routing and forwarding is performed by routing and caching mechanism. A security mechanism is used in each step of operation and mobility is supported intrinsically if the client or the host is mobile.
\\[5pt]
ICN presents some advantages when compared with the current system such as scalable and cost-efficient data distribution, persistent and unique naming, mobility support, built-in security and disruption tolerance.\cite{ICNVideoPart1} However, the mechanism that is needed to be built to provide some of these advantages has not been solved conveniently and fully up until now. For example, various naming mechanisms have been proposed to provide persistent and unique naming including flat or hierarchical names and hashes or human-readable strings representing data-objects as well as routers but each gives rise to new challenges and problems that affect the applicability of ICN.\cite{ICNSurvey2014} Because each subtopic of the architecture strictly depends on the realization of others, it is needed to be studied individually, as if the rest of the subtopics has been resolved. In this project, we decided to work on name resolution and routing part of the ICN by assuming that the name resolution problem is already solved.

\subsection{Overview of Routing in the ICN}
In order to describe the routing approaches in ICN, currently and commonly used routing algorithms must be mentioned; Link State Routing algorithm and Distance Vector Routing algorithm. By the help of these algorithms as well as OSPF specifications, routers can gather knowledge about network topology and content locations and determine the forwarding path of content in a fast and efficient way. While inspiring from operations of these algorithms, Hemmati and Aceves have described new approaches to ICN routing which are explained below.\cite{Hemmati2015}\cite{Aceves2014}
\\[5pt]
Both Link-State Content Routing(LSCR) and Distance-Based Content Routing(DCR) algorithms are proposed to be used as routing algorithms in ICN. In both of them, it is assumed that data/content is kept in the caches of multiple routers which are distributed over the network and generally closer to clients than regular servers. Information about the data/content is kept in a track by using its content as a base, via data prefixes. Both algorithms discover any content and its location by prefix search. Routing tables keep specific information about those prefixes to use them in forwarding. The major aim of these algorithms is to gather content information and to determine the closest path to content in the network layer. In some concrete version of ICN, information gathering is done on the application layer, which brings extra burden on the application layer. These algorithms put information gathering procedure on the network layer where it belongs considering the aim of the ICN. 
\newpage

\section{State of The Art}
The main consideration of this project is to prevent possible congestion that might occur on the path shared by multiple routers which must reach the same \emph{King Anchor}\cite{Hemmati2015} by definition. Because of shortest path choice of LSCR protocol, network links may not be utilized overall and a particular producer may have to deal with overloading. Since LSCR does not take the load into account while selecting the producer or the link, it may introduce additional delay and jitter rather than bringing enhancement into the traditional networking. As a result, we would like to give some solution to load balancing issue to improve the employability of the LSCR protocol.

\subsection{Summary of Research Papers}
\enlargethispage{\baselineskip}
Information Centric Networking can be deemed as novice and under-worked topic when comparing the rest of the network related subjects. The oldest paper related to the ICN goes back to the beginning of 2000s where the ICN is initially proposed.\cite{ICNSurvey2014}
\\[5pt]
\begin{figure}[H]
\centering
\includegraphics[width=1.0\textwidth]{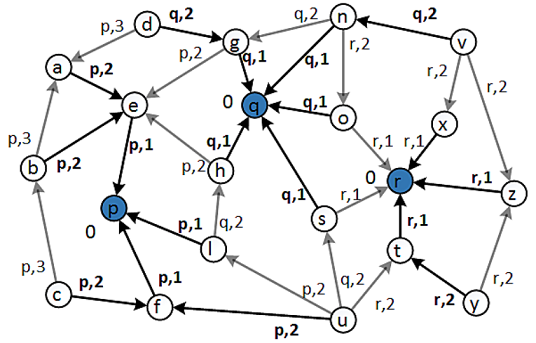}
\caption{\label{fig:linkLSCR}Basic LSCR topology representation}
\end{figure}

\newpage
One of the most wipe scope explanatory papers is the Survey of Information-Centric Networking Research conducted by G.Xylomenous and his colleagues.\cite{ICNSurvey2014} This paper explains the general view of the ICN, groups the work is done on it under five and explains what each of them is, why to work on and where to use; summaries previous researches and works from the point of five subtopics; compares and contrast previous researches and exhibits their differences. 
\\[5pt]
Link State Content Routing (LSCR) has been proposed by E.Hemmanti and J.J. Garcia-Luna-Aceves,\cite{Aceves2014} which is built on top of a common use routing algorithm, link-state routing. LSCR uses LSR functionalities and message procedures along with its new message sequences and routing tables. LSCR builds the network topology of the network like LSR does but it also builds a tree of locations where the named-data available in the network is served. It updates data locations via specific update messages different than those used for topology updates. LSCR creates multiple tables to calculate minimum distance from a router to the nearest location that the interesting data is served. Information forwarding is handled by tables and shortest path algorithms. 
\\[5pt]

\begin{figure}[H]
\centering
\includegraphics[width=1.0\textwidth]{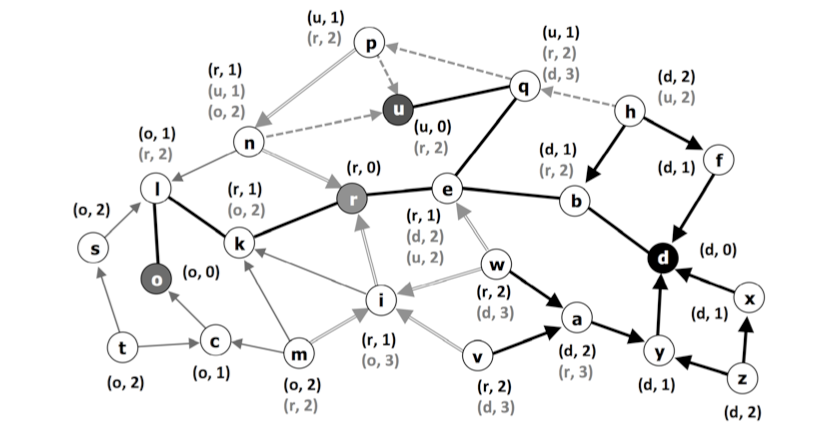}
\caption{\label{fig:vectorDCR}Basic DCR topology representation}
\end{figure}
\newpage

\subsection{Shortcomings of ICN}
ICN has been studied as 5 different approaches. These are naming, name resolution and data routing, caching, mobility and security. In these studies, our concern is only data routing \cite{ICNSurvey2014}. We accept that other approaches were already solved or will be eventually solved in future. For implementing data routing properly, routers need to keep prefixes and each prefix leads to a set of data sources/metadata information. Routers send data by the help of these prefixes and we accept that prefix naming problem are already solved. Also, we accept that we can cache any prefix (actual data object) at any router. Security and mobility is not our concern at this point.
\\[5pt]
One of the biggest handicaps of the proposed ICN architecture is the massive hash tables that may be built at the end of the information gathering process. Since networks can be narrowed and grouped as small autonomous systems, routing tables are not expected to grow over. (1000-10000 entries maybe). Forwarding tables, on the other hand, have the potential to grow drastically due to the enormous naming space of prefixes/data. Fortunately, most routers have enough internal memory space but namespaces still can go crazily big to outgrow memory space of any router. 
\newpage

\section{Methods}
\subsection{ns3 Library}
ns-3 is a discrete-event network simulator, targeted primarily for research and educational use. It is a free software having been developed since the 80s, publicly available, well designed and documented. The goal of the ns-3 project is to develop a preferred, open simulation environment for networking research. The library includes all different network specifications, almost all routing algorithms along with IPv4 and IPv6 standards. It's also fully capable of all kinds of discrete event simulations related to networking. 
\\[5pt]
Unfortunately, ns-3 is not compatible with either the named-data networking or information-centric networking. We were able to implement a kind of layer which can be seen as an ICN network on above but runs on IPv4 totally in the bottom layers. However, this kind of implementation would be troublesome and would take us away from our aim. That's why we did not use this library.  

\subsection{ndnSIM Module }
ndnSIM has been developed for researches of named-data networking on the top of the ns-3 library. It uses all the facilities of the library and almost fully implements the ICN architecture protocol stacks. ndnSIM module was implemented concerning following \href{https://named-data.net/project/specifications/}{named-data networking packet specifications.}. Although it is nice to have specifications and documentation this approach tied out hands because it did not allow us to use the new protocol stack with the old one.
\\[5pt]
From many perspectives, it would be a nice tool to implement our routing algorithm and visualize our results. However, unlikely ns-3, ndnSIM tool did not allow us to implement our routing algorithm since it is built on top of a really specific structure. Besides, we realized that ndnSIM seldom uses ns-3's core IP protocols and only makes benefit from ns-3's core event-based simulator. Thus, although it is one of the rare tools to be used in named-data networking we are unable to implement our algorithm in ndnSIM.
\newpage

\subsection{Simulator Implementation}
We wanted a flexible and mix structure to test the effect of multi-source, multi-path routing. Our main goal is not to build a fancy and generic network simulation tool but to be able to test the differences between the two methods. Therefore, the Java-based simulator is designed as simple as possible. 
\\[5pt]
ICN architecture is implemented inside the network nodes. Forward interest base and routing interest base are implemented as described in LSCR paper \cite{Hemmati2015}. Data objects are abstracted by Data classes but given access through Prefix class which represents meta-information about data. We preferred to relate each data object with each prefix, one-to-one relation rather than one-to-many or many-to-many for ease of use. Pending interest base, which is another table explained in LSCR paper \cite{Hemmati2015} is ignored because of two reasons: first, this table is designed to enhance caching mechanism which is out of our topic and second no matter which method we simulate, it will work the same.
\\[5pt]
The network is built as simple as possible. It only consists of nodes, links and packets. We consider two types of packets: interest and data packets and assigned them standard size 0.1 MB and 8 MB respectively. Links were full-duplex and had their buffers for each direction. Link capacities were assigned randomly from 512 Mbps to 2048 Mbps. All messages and packets stem from OSPF are ignored and dropped. Since it does not matter whichever underlying protocol stack applied as long as they are compatible both methods will affect the performance likewise.
\\[5pt]
We designed the simulator as a discrete event simulator. Events were either a send/receive event or an initialize interest/data event. Send/receive events are for regular packet send and receive events were as interest/data events were used to split requests for prefixes. The initialize interest event created multiple send packets based on data size and assigned paths based on which routing type was selected: single (1 Path) or multiple (3 Path). Then the event put those send packets in the respective send buffer (link from this to next hop) in the respected node (consumer). The initialize data event on the other hand was triggered after receiving an interest packet. The event then reversed the path, created a send data event and put it on the respective buffer. 
\\[5pt]
We chose source routing as the default routing type. Although we did not test the methods according to source routing and LSCR routing we thought source routing is more suitable to ICN architecture. We chose cost update function as common 1/(capacity-load) where load refers to the amount of data flow per second. In each path update (k-path Dijkstra) event, cost update function was called and then paths were recalculated. We chose the frequency of path update event as 5 updates per second as common.
\\[5pt]
We created each topology and each event sequence randomly. In the main function, simulator got running parameters (name of input and output files) then built random topology with 10 nodes, 30 edges and 15 prefixes. A particular number of (1000,5000,10000,20000) initialize interest events were made for each random node, random prefix and random time. Initialize interest events were kept in the global queue with other send/receive events. Simulator necessarily started running by the time it gets the earliest initialize interest event. 
\\[5pt]
Outputs of the simulator (load logs and packet logs) were printed as CSV file and analyzed by using some scripts and data visualization tools (pandas, NumPy, matplotlib, seaborn etc.). A script run the simulator for particular configurations for once or for multiple times and constructed stats and graphs of the particular configuration. We specified evaluation metrics like standard deviation of link loads per second and average packet delivery time, packet delivery time refers to the time between a packet was created and terminated.
\newpage

\section{Results}
The simulator was finished but it is open to some alteration and improvement. If you would like to see development tree and source code or contribute please check the following open repository. \href{https://github.com/effervescent-shot/SeniorDesignProject}{SeniorDesignProject}\cite{SourceCode}\cite{martins2003new}

\begin{figure}[H]
\centering
\includegraphics[width=1.0\textwidth]{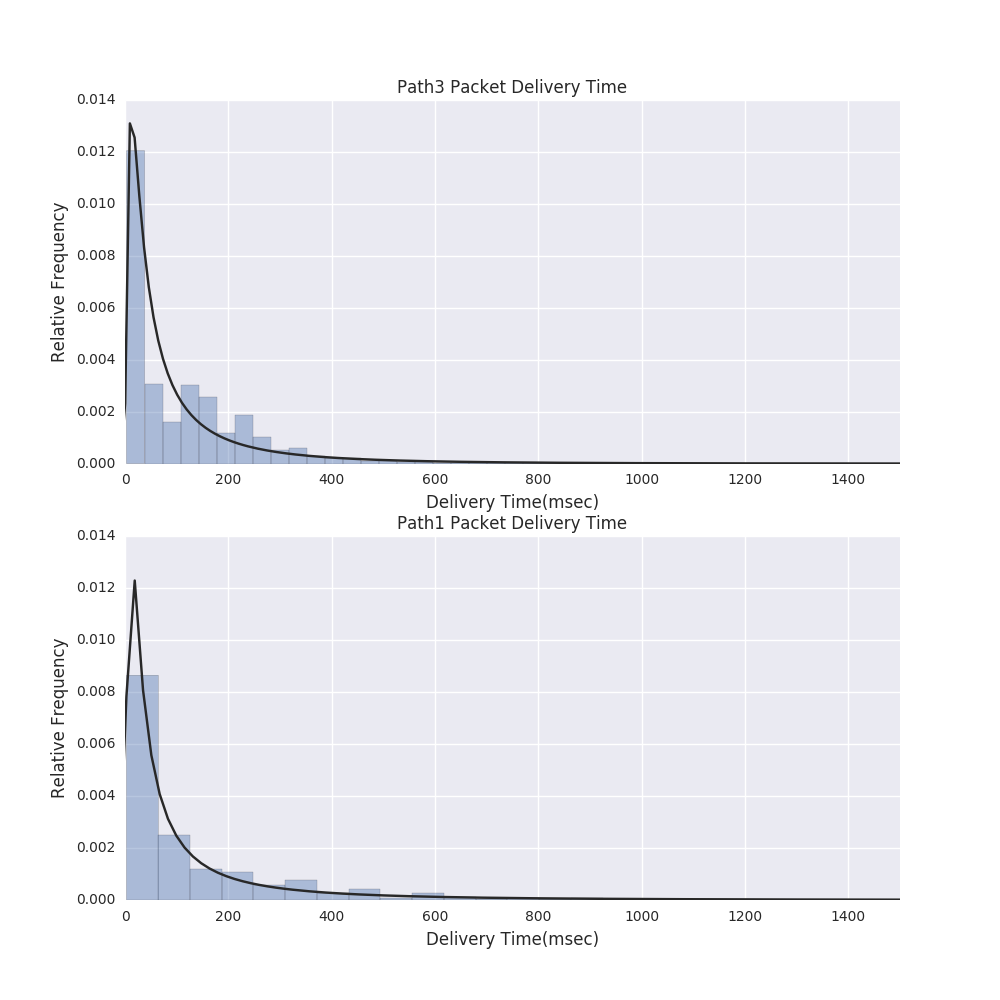}
\caption{\label{fig:singHist}Histogram of packet delivery time (single run)}
\end{figure}

The histogram shows that multiple-path routing reduces the average delivery time and also routes higher percentage of packets with smaller time.

\begin{figure}[H]
\centering
\includegraphics[width=1.0\textwidth]{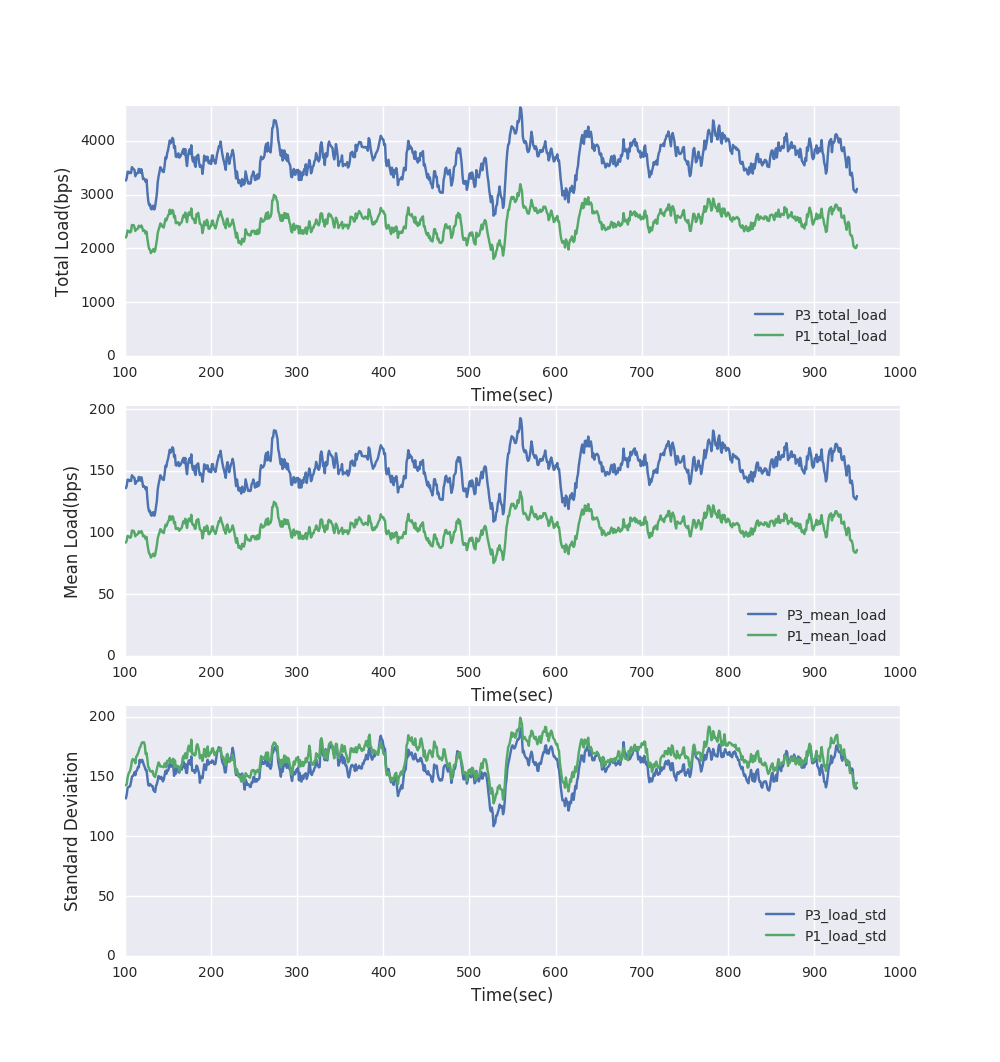}
\caption{\label{fig:singleStats} Stats of link load per second}
\end{figure}

The figured was constructed by applying convolution to time-based data and excluding warm-up (0-50s) and cool-down periods (950-1000s). It shows that multi-path routing increases the offering load as well as average load in the network. It does it so by utilizing the unused network links and closing up the load between busy link and idle link (smaller standard deviation).

\begin{figure}[H]
\centering
\includegraphics[width=1.0\textwidth]{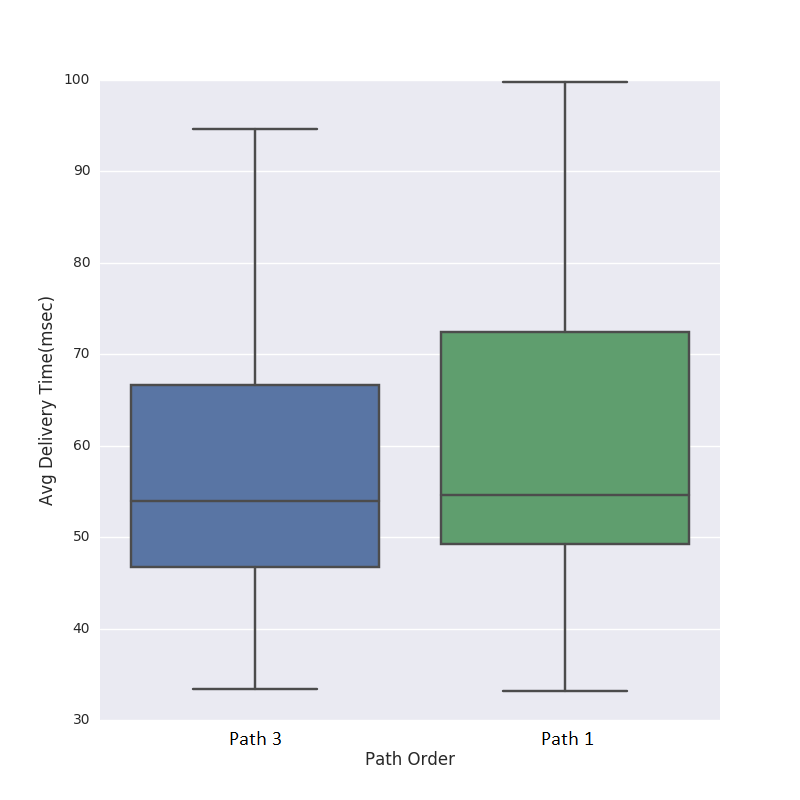}
\caption{\label{fig:multiBox}Boxplot of average delivery times (multiple run)}
\end{figure}

We expected to observe more separated distribution but still, this box plot implies that multi-path routing reduces the average delivery time regardless of the network topology and event times. 
\newpage

\section {Conclusion and Discussion}
Our graphs and statistics can be analyzed to get a proper conclusion. First, multi-source and multi-path routing provides a way to deliver a high percentage of interest/data packets faster than regular methods. Second, the new method utilizes the network better. Even for the worst-case scenarios overall offered load is higher and the standard deviation is smaller. Moreover, dynamic load awareness ensures that even under high traffic network can avoid congestions and choose the fastest, emptier path.
\\ \\
What may be better to discuss here is that we did not test and compare source routing with its alternative method. Well, it is possible to implement the LSCR algorithm but we have seen that it does not have an agreed implementation. Whoever used LSCR as routing protocol in his/her research he/she implemented the protocol in his/her way. Therefore, we would like to keep this comparison out of our project and to implement this feature whenever a common implementation is agreed. 
\newpage

\section{Future Work}
\subsection{Hypothesis Test}
We explained the evaluation metrics and concluded that multi-source routing is better. However, this claim needs validation. We are going to increase the number of repetition of simulation with various kinds of configurations and randomization. Until then we can conduct a hypothesis analysis in which we can securely reject that data regarding average delivery times under single and multi-source are coming from the same distribution. If we can reach such a conclusion we can say that multi-source routing is better. Therefore, we are going to repeat the runs, collect data and conduct a hypothesis test.    

\subsection{Producer Load}
It is possible and probably reasonable to consider the service load on the producer. Since prefixes/data that ICN concerns are generally big (video, audio) they can keep producers busy very long time. When the interest messages on a particular prefix become frequent, servers normally have difficulty in responding all these messages. In real life, this may yield a preparation delay or service delay. In the upgraded version of this simulator, we may consider producer load as another parameter which may help to choose closest sources for a prefix.
\newpage

\addcontentsline{toc}{section}{References}
\bibliographystyle{plain}
\bibliography{main}

% \section{References}
% \begin{enumerate}
% \item ICN Photo Reference. \\
% \textit{https://www.slideshare.net/SAILproject/sail-regwsicnbahlgren}

% \item J.J Garcia-Luna-Aceves Ehsan Hemmati. \emph{A New Approach to Name-Based Link-State Routing for Information-Centric Networks}. 2nd ACM Conference on ICN
% 2015 (2015), pp. 29–38 (cit. on pp. 4, 7, 8).

% \item Future Internet Architecture - ICN and CCN Part 1 (Basic part 3/5). Feb. 2017. \textit{https://www.youtube.com/watch?v=WiM7XknFWss}

% \item J.J Garcia-Luna-Aceves. \emph{Name-Based Content Routing in Information Centric Networks Using Distance Information}. 1st ACM Conference on ICN 2014 (2014), pp. 7–16 (cit. on pp. 4, 6, 8).

% \item George Xylomenos Christopher N. Ververidis Vasilios A. Siris Nikos Fotiou Christos Tsilopoulos Xenofon Vasilakos Konstantinos V. Katsaros and George C. Polyzos. \emph{A Survey of Information-Centric Networking Research} in Communications Surveys and Tutorials 2 (2014), pp. 1024–1049 (cit. on pp. 4, 6).

% \item ICN Simulator Github \textit{https://github.com/effervescent-shot/SeniorDesignProject}

% \item Yens Algorithm GitHub \textit{https://github.com/yan-qi/k-shortest-paths-java-version}

% \end{enumerate}
\end{document}